\documentclass[10pt,a4paper]{amsart}
\usepackage{amsmath}
\usepackage{amssymb}
\usepackage{amsxtra}
\usepackage{graphicx,color}
\usepackage[colorlinks]{hyperref}
\begin{document}
\sloppypar \sloppy
\newcommand{\B}{\mbox{$\mathbb{Z}_2$}} 
\newcommand{\Z}{\mbox{$\mathbb{Z}$}}
\newcommand{\N}{\mbox{$\mathbb{N}$}}
\newcommand{\R}{\mbox{$\mathbb{R}$}}
\newcommand{\C}{\mbox{$\mathbb{C}$}}
\newcommand{\pd}{\mbox{$\partial$}}
\renewcommand{\a}{\mbox{$\hat{a}$}} 
\newcommand{\p}{\mbox{$\hat{p}$}} 
\newcommand{\q}{\mbox{$\hat{q}$}} 
\newcommand{\h}{\mbox{$\hat{h}$}} 

\renewcommand{\H}{\mbox{$\mathcal{H}$}} 
\renewcommand{\L}{\mbox{$\mathcal{L}^2$}} 
\renewcommand{\O}{\mbox{$\hat{\mathcal{O}}$}} 
\newcommand{\V}{\mbox{$\mathcal{V}$}} 

\newtheorem{theorem}{Theorem}

\author{I. Schmelzer}
\thanks{Berlin, Germany}
\email{ilja.schmelzer@gmail.com}%
\urladdr{ilja-schmelzer.de}

\title{Why the Hamilton operator alone is not enough} \sloppypar

\maketitle
\begin{abstract}
In the many worlds community there seems to exist a belief that the physics of quantum theory is completely defined by it's Hamilton operator given in an abstract Hilbert space, especially that the position basis may be derived from it as preferred using decoherence techniques.

We show, by an explicit example of non-uniqueness, taken from the theory of the KdV equation, that the Hamilton operator alone is not sufficient to fix the physics. We need the canonical operators \p, \q\/ as well. As a consequence, it is not possible to derive a ``preferred basis'' from the Hamilton operator alone, without postulating some additional structure like a ``decomposition into systems''. We argue that this makes such a derivation useless for fundamental physics.
\end{abstract}

\section{Introduction}

Some quotes in the many worlds literature suggest a belief that one can derive the canonical structure from the Hamilton operator taken alone, given as an abstract linear operator in some Hilbert space, without any additional structure. For example, Tegmark describes the construction of a ``preferred basis'' in many worlds:
\begin{quote}
``This elegant mechanism is now well-understood and rather uncontroversial [\ldots]. Essentially, the position basis gets singled out by the dynamics because the field equations of physics are local in this basis, not in any other basis.'' \cite{Tegmark}
\end{quote}
This is (as indicated by the ``essentially'') an oversimplification: The decoherence-based construction considered there depends not only on the dynamics (the Hamilton operator), but also on some ``subdivision into systems'' --- a tensor product structure --- as can easily be seen in the quoted papers by Zurek:
\begin{quote}
``One more axiom should [be] added to postulates (i) - (v): (o) The Universe consists of systems.'' \cite{Zurek2007}
\end{quote}
But some comments made by Zurek suggest that he shares the belief that physics is completely defined by the Hamilton operator as well:
\begin{quote}
``Both the formulation of the measurement problem and its resolution through the appeal to decoherence require a Universe split into systems. Yet, it is far from clear how one can define systems given an overall Hilbert space of everything and the total Hamiltonian.'' \cite{Zurek1998}

``[A] compelling explanation of what are the systems --- how to define them given, say, the overall Hamiltonian in some suitably large Hilbert space --- would be undoubtedly most useful.'' \cite{Zurek1998}
\end{quote}
Indeed, the problem ``how'' to define these systems seems to assume, at least implicitly, that these systems can be defined given the Hamiltonian. Then, based on the decoherence technique, the preferred basis can be defined as well.

The following quote suggests that Vaidman shares this belief too:
\begin{quote}
I believe that the decomposition of the Universe into sensible worlds \ldots is, essentially, unique. The decomposition, clearly, might differ due to coarse or fine graining, but to have essentially different decompositions would mean having a multi-meaning Escher-type picture of the whole Universe continuously evolving in time.
\cite{Vaidman}
\end{quote}
If we interpret the ``decomposition into sensible worlds'' as something based on the decoherence-constructed ``preferred basis'', the uniqueness of this decomposition implies the uniqueness of the ``preferred basis'' as well. 
Schlosshauer talks about
\begin{quote}
the physical definition of the preferred basis derived from the structure of the unmodified Hamiltonian as suggested by environment-induced selection \cite{Schlosshauer}
\end{quote}
which also suggests that he shares the belief. Last but not least, let's quote Brown and Wallace \cite{Wallace}. They discuss the possible non-uniqueness of the ``preferred basis'' picked out by decoherence:
\begin{quote}
``\ldots granted that decoherence picks out a quasi-classical basis as preferred, what is to say that it does not also pick out a multitude of other bases -- very alien with respect to the bases with which we ordinarily work, perhaps, but just as 'preferred' from the decoherence viewpoint. Such a discovery would seem to undermine the objectivity of Everettian branching, leaving room for the Bohmian corpuscle to restore that objectivity.'' \cite{Wallace}
\end{quote}
and present the following response as preferable:
\begin{quote}
``Granted that we cannot rule out the possibility that there might be alternative decompositions, and that this would radically affect the viability of the Everett interpretation -- well, right now we have no reason at all to suppose that there actually are such decompositions. Analogously, logically we can't absolutely rule out the possibility that there's a completely different way of construing the meaning of all English words, such that they mostly mean completely different things but such that speakers of English still (mostly) make true and relevant utterances. Such a discovery would radically transform linguistics and philosophy, but we don't have any reason to think it will actually happen, and we have much reason to suppose that it will not. To discover one sort of higher-level structure in microphysics (be it the microphysics of sound-waves or the micro-physics of the wave-function) is pretty remarkable; to discover several incompatible structures in the same bit of microphysics would verge on the miraculous.''  \cite{Wallace}
\end{quote}

The aim of this paper is to show that this miracle happens --- the theory of the Korteweg - de Vries (KdV) equation gives nice counterexamples for this thesis: If a potential $V(q,s)$ is a solution of the KdV equation, then the operators $\h(s)=-\pd_q^2 + V(q,s)$ for different $s$ appear to be unitarily equivalent, despite defining different physics. Thus, the physics of canonical quantum theories is not completely defined by the Hamilton operator alone.

This fact seems fatal for the idea to derive a preferred basis, using decoherence techniques, from the Hamilton operator taken alone. One needs an additional structure --- be it the tensor product structure related with the ``decomposition into systems'' or whatever else --- which has to be postulated. We consider the question if this construction could be, nonetheless, used as a foundation of quantum theory, as a replacement for simply postulating the configuration space. We argue that this construction of the preferred basis combines the disadvantages of postulated structures (lack of explanatory power) and emergent structures (uncertainty, dependence from other structures, especially dynamics), and, therefore, should be rejected in favour of the canonical way to postulate the configuration space as a non-dynamical structure, as done in canonical quantum theories as well as pilot wave theories.

Thus, the ``derivation'' of the ``preferred basis'' based on decoherence techniques seems useless in the domain of fundamental physics. It has it's useful applications in situations where we already have (as in the Copenhagen interpretation or in pilot wave theories) a classical part, which defines the decomposition into systems which one needs to derive a decoherence-preferred basis. 

\section{Why one can think that the Hamilton operator is sufficient}

Some clarification of how we interpret this belief seems useful. 

Canonical quantum theories are defined in a more or less standard way: First, one defines the kinematics by defining some Hilbert space \H\/ with some set of canonical operators $\p_i$, $\q^i$ with commutation relation $[\p_i,\q^j] = -i\hbar\delta^j_i$ on \H, or, equivalently, to postulate some configuration space $Q$ with coordinates $q^i$ so that $\H\cong\L(Q,\C)$ and $p^i=i\hbar\pd_i$. Then, in a second step, one defines the dynamics by postulating the Schr\"{o}dinger equation
\begin{equation}
 i\hbar\pd_t\psi(t) = \h \psi(t),\qquad \psi \in \H
\end{equation}
for some Hamilton operator \h, usually of the form
\begin{equation}\label{HinQ}
 \h = \h(\p_i,\q^i) = \sum_i \frac{1}{2m_i}\p_i^2 + V(\q^i).
\end{equation}
Implicit part of the definition of the canonical operators $\p_i$, $\q^i$ is their identification with classical observables. It is hard to press this part of the definition into some formal property --- it depends on the particular theory. The point is that the definition is only complete if we know what it means to measure the configuration $Q$ --- it is some procedure, usually known from the corresponding classical theory. The details of this procedure do not matter in a general discussion. But in order to apply the theory to make concrete predictions, one has to know how to measure the $q_i$ and $p^i$. Without this additional information the physical definition of the theory is not complete. The quantum theory predicts the result of this measurement for a state $\psi(q)\in\L(Q,\C)$ as $|\psi(q)|^2$. But this information would be useless if we don't know how to measure $Q$.

Given this physical meaning of $Q$, it seems, at a first look, completely unreasonable to think that the physics is completely defined by the Hamilton operator alone. Nobody would apply, for example, some unitary transformation $U$ to $\h\to U\h U^{-1}$ and $\psi\to U\psi$, but not to $\p_i$, and $\q^i$, but nonetheless claim that the physics remain unchanged. Everybody knows that one has to apply the same unitary transformation to $\p_i\to U\p_i U^{-1}$ and $\q^i\to U\q^i U^{-1}$ as well to preserve physics.

Now, with our interpretation of the quotes, we are in no way suggesting that the authors do not recognize this. The idea that the Hamilton operator alone is sufficient is a different one. Given the very special form of the Hamilton operator \eqref{HinQ}, one can imagine that it may be possible to \emph{reconstruct} the configuration space $Q$ more or less uniquely given \h\/ as an abstract operator on some abstract Hilbert space \H. Last but not least, the straightforward ideas to construct counterexamples fail: One can apply coordinate transformations $q^{i'} = q^{i'}(q^i)$, but these do not change the physics, and leave the abstract configuration space $Q$ unchanged. One can consider a canonical rotation $\p\to \q$, $\q\to-\p$, but this gives a very strange nonlocal operator of type $\h = \q^2 + V(i\pd_q)$, which forbids a physical interpretation of the new $\q$ in terms of a configuration. Therefore, one can conclude that there are only few sufficiently nice choices of $\q^i$ which allow a meaningful interpretation of the $\q^i$ as configuration space coordinates, and one could hope or expect that these few choices lead, at least approximately, to physically equivalent theories. In this case, it would be indeed the Hamilton operator taken alone, as an abstract operator \h\/ on a Hilbert space \H, which would be sufficient to define the physics completely.

But this expectation is false, as we will show below.

\section{What we can learn from the Korteweg - de Vries equation}

A really beautiful one-dimensional example of different representations follows from the theory of solutions of the Korteweg-de Vries (KdV) equation:
\begin{theorem}
If the function $V(q,s)$ is a solution of the Korteweg-de Vries equation
\begin{equation}\label{eq:KdV}
 \pd_s V(q,s) = - \pd_q^3 V(q,s) + 6 V(q,s) \pd_q V(q,s),
\end{equation}
then the operators
\begin{equation}\label{hs}
 \h(s) = -\pd_q^2 + V(q,s)
\end{equation}
for different $s$ are unitarily equivalent. 
\end{theorem}
Indeed, as has been found by Lax (\cite{kdv-Lax}), the KdV equation is equivalent to the operator equation
\begin{equation}\label{eq:Lax}
 i\pd_s \h(s) = [\a(s),\h(s)].
\end{equation}
for the self-adjoint operator
\begin{equation}
 \a(s) = i(-4\pd_q^3 + 6 V(q,s)\pd_q + 3 (\pd_q V(q,s))),
\end{equation}
as one can easily check. But this type of operator equations defines, for a self-adjoint operator $\a(s)$, a unitary evolution of $\h(s)$. Indeed, this is simply the analogon of the Heisenberg equation for the ``Hamilton operator'' $\a(s)$, applied to $\h(s)$. The unitary transformation we need is defined by the equation
\begin{equation}
i\pd_s U(s) = \a(s)U(s), \qquad U(0) = 1.
\end{equation}

For results about the existence of solutions of the KdV equation see, for example, \cite{kdv-Kametaka}.

This representation leaves \p\/ and \q\/ invariant but changes $\h(s)$ and $\psi(t,s)$ by $\h(s)=U(s)\h(0)U(s)^{-1}$ resp. $\psi(t,s)=U(s)\psi(t,0)$. This is not yet exactly our point. But, given the unitary operators $U(s)$, we can also consider another equivalent representation:

\begin{theorem}\label{hfixed}
For a given Hamilton operator $\h = -\pd_q^2 + V(q,0)$, with $V$ as defined by \eqref{eq:KdV}, there exist canonical operators $\q(s)$, $\p(s)$, so that the representation of \h in terms of $\q(s), \p(s)$ is given by \eqref{hs}.
\end{theorem}

Indeed, it is sufficient to define the operators $\q(s)$, $\p(s)$ by $\q(s)=U(s)^{-1}\q U(s)$ and $\p(s)=U(s)^{-1}\p U(s)$. In this case, the unitary transformation $\hat{O} \to U(s)\hat{O}U(s)^{-1}$, applied to the operators $\hat{O} \in\{\h, \q(s), \p(s)\}$, gives $\{\h(s), \q, \p\}$, thus, the representation of \h\/ in terms of $\q(s), \p(s)$ is $\h(s)$.

It is worth to note that the Hamiltonian \h\/ in this theorem does not depend on $s$. Only the operators $\q(s), \p(s)$ depend on $s$. This has the consequence that the representation $\h = \h(\p(s), \q(s))$ in terms of these $s$-depending operators depends on $s$ too. But \h\/ is, nonetheless, the same for all $s$.

Thus, given the operator \h\/ alone, we cannot reconstruct the operators \p, \q\/ uniquely. The operators $\p(s)$, $\q(s)$ give equally nice candidates for canonical variables for \h: The representation of \h\/ in these operators has the same canonical form $\h = -\pd_q^2 + V(q)$, and the potential functions $V(q)=V(q,s)$ in this standard form are as nice and well-behaved as the original $V(q)=V(q,0)$.

\section{Different canonical structures define different physics}

\begin{figure}
\centerline{\includegraphics[angle=0,width=0.8\textwidth]{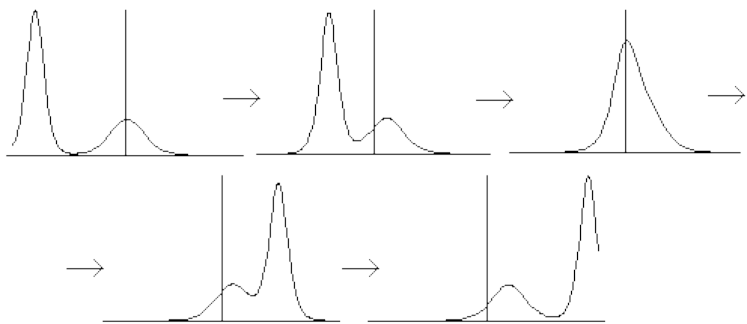}}
\caption{\label{fig:solitons}
Two-soliton-solution $u(q)=-V(q,s)$ of the KdV equation for different values of the evolution parameter $s$. Picture taken from \cite{Takasaki}}
\end{figure}

Let's clarify if the different potentials $V(\q,s)$ really define different physics. This seems obvious, if one looks at particular examples, like in fig. \ref{fig:solitons}. The two sharply localized part of the solutions in the first and last picture of figure \ref{fig:solitons} are so-called solitons, special solutions, which, taken alone, have the exact form
\begin{equation}
V(q) = -\frac{\lambda}{2}\cosh^{-2}\left(\frac{1}{2}\sqrt{\lambda}(q - \lambda s - q_0)\right)
\end{equation}
and move with velocity $\lambda$. Their spectrum is defined by a single eigenvalue $E(\lambda)$, with an eigenfunction $\psi_\lambda(q)$ localized in the same domain. Some superpositions of the two eigenstates would be, in one case (first and last picture), clearly delocalized, in another one (upper right picture) they are all localized in the same region.

\begin{figure}
\centerline{\includegraphics[angle=0,width=0.6\textwidth]{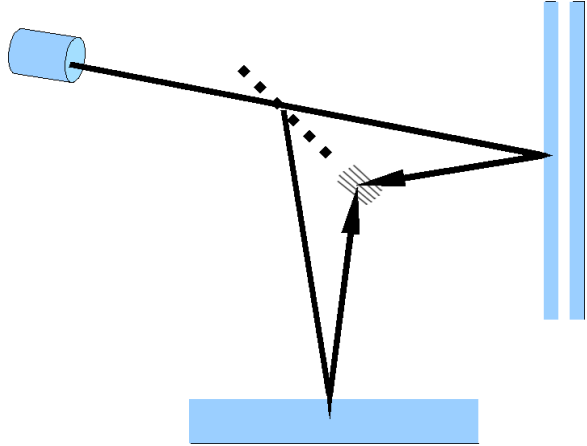}}
\caption{\label{fig:reflection} An experiment which would allow us to measure a phase difference between the reflection coefficients $b(k)$ of the horizontal and vertical mirrors. For reflection angles close enough to $90^o$, only the one-dimensional reflection coefficient in orthogonal direction matters. Thus, we can put the different one-dimensional potentials for different $s$ into the mirrors, extended trivially in the other direction, say, a localized one at the bottom, and one with two solitons on the right.}
\end{figure}

We can also consider the scattering matrix. The inverse scattering method (developed by \cite{kdv-Gardner}, see \cite{kdv-book}) for solving the KdV equation gives the following explicit result for the one-dimensional scattering matrix: One of the two coefficients of the scattering matrix, namely $a(k)$, appears to be an integral of motion of the KdV equation. Instead, the reflection coefficient $b(k)$ depends explicitly on $s$. To construct an experiment which allows the measurement of such differences is not difficult (see fig. \ref{fig:reflection}). If not a different scattering matrix, what else defines different physics?

\section{The non-uniqueness of the tensor product structure}

The example we have given is one-dimensional. One can consider a straightforward generalization to higher dimension by considering Hamilton operators of the form
\begin{equation}
 \h(s) = \sum \frac{1}{2m_i}\pd_i^2 + V_i(q^i,s_i)
\end{equation}
with different one-dimensional potentials $V_i(q^i,s_i)$. But all we obtain in this way are only non-interacting degrees of freedom. This seems to leave some hope for the case of non-trivially interacting Hamiltonians in higher dimension. The interactions between different degrees of freedom could, possibly, allow the choice of a preferred basis.

This is, essentially, the way used in the decoherence-based approach for the construction of a preferred basis. One starts with a ``decomposition into systems'':
\begin{quote}
(o) The Universe consists of systems.\cite{Zurek2007}
\end{quote}
that means, a tensor product structure
\begin{equation}\label{Hproduct}
\H\cong \H_1 \otimes \H_2 \otimes \ldots = \bigotimes_i \H_i
\end{equation}
on the Hilbert space of everything \H. In general, each factor $\H_i$ interacts with it's environment $\H_E \cong \bigotimes_{j\neq i}\H_j$. In particular, if \h\/ has the standard form
\begin{equation}\label{Hstandard}
\h = \sum \frac{1}{2m_i}\p_i^2 + V(\q)
\end{equation}
the interaction Hamiltonian is $V(\q)$. The preferred basis is, in a simplified version,
\footnote{
In general, the situation is more complex, because one has to take into account also the self-Hamiltonian $H_S$ of the system:
\begin{quote}
In the more general circumstances the states which commute with $H_{SE}$ at one instant will be rotated (into their superpositions) at a later instant with the evolution generated by the self-Hamiltonian $H_S$. \ldots An example of this situation is afforded by a harmonic oscillator, where the dynamical evolution periodically “swaps” the state vector between its position and momentum representation, and the two representations are related to each other by a Fourier transformation. In that case the states which are most immune to decoherence in the long run turn out to be the fixed points of the “map” defined by the Fourier transformation. Gaussians are the fixed points of the Fourier transformation (they remain Gaussian). Hence, coherent states which are unchanged by the Fourier transform are favored by decoherence.
\cite{Zurek1998}
\end{quote}
But for the purpose of this section --- to prove the non-uniqueness of a tensor product structure with sufficiently nice properties --- this simplified version is sufficient. Last but not least, if the nice tensor product structure is unique, nobody can forbid us to apply the simplified version to construct a unique configuration space. But this is what seems to be impossible.
}
the one which is measured by the interaction Hamiltonian. In case of $V(\q)$ as the interaction Hamiltonian between $\H_i$ and its environment $\H_E$, it is $\q^i$ which is measured by the environment.

Thus, we can recover the operator $\q^i$ on $\H_i$ by taking into account the interaction with the environment. The one-dimensional counterexample, as well as the straightforward non-interacting examples, could be interpreted as  irrelevant exceptions, which play no role in the multi-dimensional interacting case.

Unfortunately, this construction depends on the predefined tensor product structure. And this tensor product structure is not unique:

\begin{theorem}\label{th:tensorproduct}
There exists a Hamilton operator \h\/ in a Hilbert space \H\/ such that there exist different tensor product structures $\H\cong \H_{1s}\otimes\H_{2s}$ with canonical variables $\q_{1s},\p_{1s}$ and $\q_{2s},\p_{2s}$ on the factor spaces $\H_{1s}$ resp. $\H_{2s}$ so that \h\/ has in all of them the standard canonical form
\begin{equation}
 \h = \p_{1s}^2 + \p_{2s}^2 + V(\q_{1s},\q_{2s},s),
\end{equation}
with a non-trivial interaction potential $V(\q_{1s},\q_{2s},s)$, which depends non-trivially on $s$.
\end{theorem}

This can be easily seen in a minor variant of the straightforward multidimensional extension of the KdV example. Let's start with a simple degenerated two-dimensional Hamilton operator
\begin{equation}
\h = \p_x^2 + \p^2_y + V(\q^x,0) + V(\q^y,s).
\end{equation}
Similar to theorem \ref{hfixed}, we choose \h\/ as well as $\q^x,\p_x$ as fixed, but $\q^y=\q^y(s),\p_y=\p_y(s)$ as depending on $s$. Now, we define the tensor product structure we need by
\begin{equation}
\begin{split}
\q_{1s} = \frac{1}{\sqrt{2}}(\q^x + \q^y(s));\qquad  &\p_{1s} = \frac{1}{\sqrt{2}}(\p_x + \p_y(s);\\
\q_{2s} = \frac{1}{\sqrt{2}}(\q^x - \q^y(s));\qquad  &\p_{2s} = \frac{1}{\sqrt{2}}(\p_x - \p_y(s)).
\end{split}
\end{equation}
In these variables, the interaction potential is already nontrivial. But, it has yet the same nice standard canonical form, and as in theorem \ref{hfixed}, the resulting potential
\begin{equation}
 V(\q_{1s},\q_{2s},s) = V(\q^x,0) + V(\q^y,s)
\end{equation} 
is of comparable nice quality for different $s$. The tensor product structure depends on $s$. A nice implicit way to see this is to use the fact that the simplified version of decoherence already allows the unique derivation of the positions $\q_{1s},\q_{2s}$ as the decoherence-preferred observables, for the given tensor product structure $\H_{1s}\otimes\H_{2s}$. But, on the other hand, the result is obviously not unique. This contradiction disappears once we recognize that the tensor product structure is not unique. But the $s$-dependence of the tensor product structure can be seen directly as well: If the tensor product structure would be the same for different $s$, we would be able to express $\q_{1s}$ as a function $\q_{1s}=F(\q_{10},\p_{10})$. But an attempt to express $\q_{1s}$ in this way fails for a general $U(s)$ --- there is no chance to get rid of the dependence on $\q_{20}$:

\begin{equation}
\q_{1s} = \frac{1}{2}\left((\q_{10}+\q_{20}) +  U(s)^{-1}(\q_{10}-\q_{20})U(s)\right) \neq F(\q_{10},\p_{10})
\end{equation} 

Let's emphasize again that the operator \h\/ does not depend on $s$. Only the operators $\p_{is},\q_{is}$ depend on $s$. Therefore, also the representation of \h\/ in terms of these operators depends on $s$. But the operator \h\/ itself has not only the same spectrum, but is simply the same for different $s$.

As a consequence, one should give up the hope expressed by Zurek:
\begin{quote}
``[A] compelling explanation of what are the systems --- how to define them given, say, the overall Hamiltonian in some suitably large Hilbert space --- would be undoubtedly most useful.'' \cite{Zurek1998}
\end{quote}
Already in the two-dimensional case there is no unique way to reconstruct a physically reasonable, nice tensor product structure, or ``decomposition into systems'', from a given Hamilton operator taken alone.

\section{Can the decoherence-based construction replace the configuration space?}
\label{sec:Comparison}

But, maybe the decoherence-based reconstruction of the configuration space basis is, nonetheless, worth something? Last but not least, even if it depends on some other structure --- doesn't it, nonetheless, explain something important about the fundamental nature of the configuration space? We don't think so, and the aim of this section is to explain why.

Note that we do not want to question at all that there are lot's of useful non-fundamental applications of this construction --- applications where a subdivision into systems is defined by the application. The systems in these applications will be various measurement instruments and state preparation devices, various parts of the environment, and the quantum system which is interesting in the particular application. The preferred basis, constructed in this way, is also application-dependent. While it may be very important to find such a basis for a particular application, these constructions seem irrelevant in considerations of the foundations of physics.

A variant is to start from a tensor product decomposition given by the fundamental, postulated configuration space $Q\cong\prod_i Q_i$. The resulting decoherence-preferred basis may be different from the position basis and better suited for the consideration of the classical limit. But this would be a non-fundamental application as well, with no relevance for the foundations of quantum theory, which is, in this variant, postulated in the usual way.

The only variant which seems relevant for fundamental physics is to replace the a-priori definition of the configuration space in canonical quantum theories as well as pilot wave theories by the decoherence-based construction of the preferred basis, based on some fundamental tensor product structure.

To evaluate this replacement, we have to compare it with the standard alternative --- to postulate the configuration space $Q$ as a predefined, non-dynamical structure.

First, the above competitors depend on predefined, non-dynamical structures, which are introduced into the theory in an axiomatic way. The standard approach postulates the configuration space $Q$ itself, and the decoherence-based construction postulates a tensor product structure. In this sense, they are on equal footing.

But the resulting structure --- the configuration space $Q$ --- is, in the standard approach, a predefined, non-dynamical object. Instead, in the decoherence-based construction, it depends on dynamics. This dependence of $Q$ on dynamics has at least one obvious disadvantage: We can no longer define the dynamics in the canonical way as
\begin{equation}
 \h = \Delta + V(q),
\end{equation}
which uses simple and natural structures on $Q$ --- the Laplace operator $\Delta$ on $Q$, as well as the special subclass of multiplication operators in $\L(Q,\C)$, and, in addition, some special function $V(q)$ like $1/|q^1-q^2|$. Indeed, such a definition of \h\/ in terms of $Q$ would become circular. Thus, we loose a nice and simple way to define the dynamics of the theory. We would have to define the dynamics in some other way. We can expect that this other way is more complex, and less beautiful.

Thus, for the emergent, derived character of $Q$ we have to pay. In itself, this is not untypical for emergent objects: We have to pay some costs for the possibility to explain them. In a previous theory, they have been postulated as fundamental, simple, independent objects, with some nice, well-defined properties. In the new theory, which derives them as emergent objects, they become more complex, often with uncertain boundaries, and they depend on other objects and structures. Nonetheless, the special character of the loss in our case seems untypical even for emergent objects. Usually, the structures which depend on the objects which now become emergent, are or of some higher, emergent level already in the previous theory, or they become emergent together with these objects in the new theory. I don't know of an example of an object which depends on another object in the old theory, and then the other object becomes emergent, but the object itself remains fundamental, as it would be the case for \h. (Of course, this could happen, but it would indicate that the dependencies were wrong already in the old theory.) Thus, I would characterize the costs related with the dependence of $Q$ on the dynamics as higher than usual for emergent objects.

Let's consider now what we gain. Usually, if an object becomes emergent which was previously fundamental, our gain is explanatory power. Do we have such a gain in our case? Here, we have to take into account that we need another, postulated structure --- the tensor product structure --- to construct $Q$. Thus, we can ``explain'' the configuration space $Q$ only in a relative sense, in comparison with the  unexplained, postulated tensor product structure. Now, explaning one structure in terms of another may also give large explanatory power. Last but not least, in some sense all our more fundamental theories are of this type --- they explain the previously postulated objects and structures in terms of some more fundamental, but also postulated, objects and structures.

But we would talk about explanatory power only if the new, more fundamental structure has some advantages in simplicity, beauty, generality, or whatever else. What is the situation in our case? At least I cannot see any advantage. Instead, I see a lot of disadvantages:

First, we have simple examples of quantum theories which have natural configuration spaces but do not have a (similarly natural) tensor product structure: For example, finite-dimensional quantum theories with prime dimension do not have nontrivial tensor product structures. Indeed, from \eqref{Hproduct} follows $\textrm{dim} \H = \textrm{dim} \H_1\cdot\textrm{dim}\H_2\cdot\ldots$, thus the only possible tensor product structure for a space with prime dimensnion is the trivial one, which is worthless because it does not allow the start of the decoherence procedure. Then we have spaces of identical particles, which are factor spaces of a tensor product, but do not have their own natural
\footnote{
In the infinite-dimensional case one can always construct artificial tensor product structures, for example by applying Cantor's diagonal construction $\N \cong \N\times\N$ to an arbitrary basis $\psi_i$ of \H.
}
tensor product structure. Last but not least, for topologically sufficiently non-trivial manifolds $Q$, starting with $S^2$, there exists no decomposition $Q\cong Q_1\times Q_2 \times \ldots$ into factor-manifolds $Q_i$, and therefore no natural tensor product structure. But for all these Hilbert spaces we have natural configuration spaces.

Then, the tensor product structure is often less symmetric in comparison with the configuration space. The simplest example is the tensor product structure of $\L(\R^3)$ of one-particle theory, which destroys rotational symmetry. In addition, the tensor product structure based on points in field theory requires an identification of points in different time slices, if considered as fixed over time, destroying Galilean or relativistic symmetry.

These arguments seem sufficient to argue that a tensor product structure, as a fundamental object, is worse than a configuration space. Thus, an explanation of $Q$ in terms of a more fundamental tensor product structure has no explanatory power at all. The gain which we usually want to reach by deriving structures previously considered to be fundamental, namely explanatory power, cannot be reached in this approach.

Moreover, this cannot be hailed in some way, say, by deriving the tensor product structure from something else. The Hamilton operator alone is not sufficient, as shown by our counterexample. Thus, one needs some additional structure anyway. Therefore, the main line of our argumentation remains intact. The construction combines only disadvantages: the lack of explanatory power of postulated objects, with the uncertainty and dependency of emergent objects. What remains intact is also the dependence of $Q$ on the dynamics, and therefore the very special loss of the possibility to define the dynamics as $\Delta + V(q)$ on $Q$. What also remains is the simplicity and the general and very natural character of postulating a configuration space $Q$. Therefore, whatever the new additional structure, we cannot expect a large gain in explanatory power.

Given this situation, there seem to be no gains but only losses, in a construction which constructs the configuration space $Q$ postulated in canonical quantum theories and pilot wave theories using decoherence techniques, starting from a fundamental tensor product structure (or some replacement).

\section{Discussion}

We have shown that it is not the Hamilton operator alone which defines the physics of quantum theories. In addition, one needs the canonical configuration space $Q$, or some similar structure, which connects this Hamilton operator with observable configurations.

As a consequence, hopes to derive the configuration space basis using decoherence techniques from a Hamilton operator taken alone have to be given up. Such constructions necessarily depend on some additional structure, like a ``subdivision into systems'', which have to be postulated. The derivation of the configuration space basis from such additional structures combines the disadvantages of predefined structures (lack of explanatory power) and emergent structures (dependence on dynamics) without giving any advantages thus loses in comparison with a simply postulated configuration space.

If these arguments against a decoherence-based construction of the preferred basis are of any relevance for Everettians is a completely different question. They may also completely ignore the non-uniqueness and follow some reasoning like this:
\begin{quote}
Suppose that there were several such decompositions, each supporting information-processing systems. Then the fact that we observe one rather than another is a fact of purely local signiﬁcance: we happen to be information-processing systems in one set of decoherent histories rather than another.
\cite{Wallace}
\end{quote}
Indeed, once one introduces many worlds anyway, some more of them do not matter anymore. All what I can say about this is to recommend a further ``improvement'' in this direction --- to assign reality not only to the state vector of our multiverse, but to all other states as well, and to be consequent, to all Hamilton operators as well.
 
Another possibility would be to throw away the ``solution of the preferred basis problem'' and instead to use the same predefined configuration space as used in canonical quantization and pilot wave theories. Given the arguments in section \ref{sec:Comparison}, this would be an improvement.

For pilot wave interpretations \cite{deBroglie,Bohm}, the clarification that the configuration space is necessary to fix the physics of a canonical quantum theory is clearly helpful. It weakens a quite common argument that the choice of the configuration space in pilot wave interpretations is artificial, like
\begin{quote}
\ldots the artificial asymmetry introduced in the treatment of the two variables of a canonically conjugated pair
characterizes this form of theory as artificial metaphysics. (\cite{Pauli}, as quoted by \cite{Freire}),

``\ldots the Bohmian corpuscle picks out by fiat a preferred basis (position) \ldots'' \cite{Wallace}
\end{quote}

Instead, recognizing that the configuration space is part of the definition of the physics gives more power to an old argument in favour of the pilot wave approach, made already by de Broglie at the Solvay conference 1927:
\begin{quote}
``It seems a little paradoxical to construct a configuration space with the coordinates of points which do not exist.'' \cite{deBroglie}.
\end{quote}

\section{Acknowledgements}

Thanks Ch. Roth for correcting my poor English.


\begin{thebibliography}{99}

\bibitem{kdv-book} Ablowitz, M. J., Clarkson, P. A.: Solitons, nonlinear evolution equations and inverse scattering, London Mathematical Society Lecture Note Series, 149, Cambridge University Press, Cambridge (1991)

\bibitem{deBroglie} de Broglie, L., La nouvelle dynamique des quanta, in “Electrons et Photons: Rapports et Discussions du Cinquieme Conseil de Physique”, ed. J. Bordet, Gauthier-Villars, Paris, 105-132 (1928), English translation in: Bacciagaluppi, G., Valentini, A.: “Quantum Theory at the Crossroads: Reconsidering the 1927 Solvay Conference”, Cambridge University Press, and \href{http://arxiv.org/abs/arXiv:quant-ph/0609184}{arXiv:quant-ph/0609184} (2006)

\bibitem{Bohm}
Bohm, D: A suggested interpretation of the quantum theory in terms of ``hidden'' variables, Phys. Rev. 85, 166-193 (1952)

\bibitem{Freire} Freire Jr., O.: Science and exile: David Bohm, the hot times of the Cold War, and his struggle for a new interpretation of quantum mechanics, Historical Studies on the Physical and Biological Sciences 36(1), 1-34, 
\href{http://arxiv.org/abs/arXiv:quant-ph/0508184}{arXiv:quant-ph/0508184} (2005)

\bibitem{kdv-Gardner} Gardner, C. S., Greene, J. M., Kruskal,  M. D., Miura, R. M.: Method for solving the Korteweg–de Vries equation, Phys. Rev. Lett., 19, 1095 (1967)

\bibitem{kdv-Kametaka} Kametaka, Y.: Korteweg-de Vries equation. I. Global existence of smooth solutions. Proc. Japan Acad., 45, 552-555 (1969).

\bibitem{kdv-Lax} Lax, P. D.: Integrals of nonlinear equations of evolution and solitary waves, Comm. Pure Appl. Math., 21, 467–490 (1968)

\bibitem{Pauli} Wolfgang Pauli, Remarques sur le problème des paramètres cachés dans la mécanique quantique et sur la théorie de l’onde pilote, in André George, ed., Louis de Broglie – physicien et penseur (Paris, 1953), 33-42

\bibitem{Schlosshauer} Schlosshauer, M.: Decoherence, the measurement problem, and interpretations of quantum mechanics,  Rev. Mod. Phys. 76, 1267-1305 (2004)  \href{http://arxiv.org/abs/arXiv:quant-ph/0312059}{arXiv:quant-ph/0312059}

\bibitem{Takasaki} Takasaki, K.: Many Faces of Solitons, \hfill \phantom{.} \href{http://www.math.h.kyoto-u.ac.jp/~takasaki/soliton-lab/gallery/solitons/index-e.html}{\mbox{www.math.h.kyoto-u.ac.jp/$\sim$takasaki/soliton-lab/gallery/solitons}}

\bibitem{Tegmark} M. Tegmark, Many worlds or many words, Fortschr. Phys. 46, 855, \href{http://arxiv.org/abs/arXiv:quant-ph/9709032}{arXiv:quant-ph/9709032}  (1997)

\bibitem{Vaidman} Vaidman, L.: On schizophrenic experiences of the neutron or why we should believe in the many-worlds interpretation of quantum theory, International Studies in Philosophy of Science 12, 245-261
 \href{http://arxiv.org/abs/arXiv:quant-ph/9609006}{arXiv:quant-ph/9609006} (1998)

\bibitem{Wallace} Brown, H.R., Wallace, D.: Solving the measurement problem: de Broglie-Bohm loses out to Everett,  Foundations of Physics, Vol. 35, No. 4, 517 (2005) \href{http://arxiv.org/abs/arXiv:quant-ph/0403094}{arXiv:quant-ph/0403094}

\bibitem{Zurek1998}  Zurek, W.H.: Decoherence, einselection, and the existential interpretation, Philos. Trans. R. Soc. London, Ser. A 356, 1793-1821, \href{http://arxiv.org/abs/arXiv:quant-ph/9805065}{arXiv:quant-ph/9805065} (1998)

\bibitem{Zurek2007} Zurek, W.H.: Relative states and the environment: einselection, envariance, quantum Darwinism, and the existential interpretation, \href{http://arxiv.org/abs/arXiv:0707.2832}{arXiv:0707.2832} and Los Alamos preprint LAUR 07-4568 (2007)

\end{thebibliography}
\end{document}